\documentclass{article}
\usepackage{tikz}
\usepackage[utf8]{inputenc}
\usepackage[T1]{fontenc}
\usepackage{lmodern}
\usepackage{amsmath}
\usepackage{amssymb}
\usepackage{graphicx}
\graphicspath{{pictures/}}
\usepackage{booktabs}

\usepackage[]{xcolor}
\newcommand{\colorpalette}[1]{%
    \colorlet{#1l}{#1!50!white}%
    \colorlet{#1ll}{#1!20!white}%
    \colorlet{#1lll}{#1!10!white}%
    \colorlet{#1d}{#1!90!black}%
    \colorlet{#1dd}{#1!70!black}%
}
\definecolor{colorf}{HTML}{734f96}%
\colorpalette{colorf}
\definecolor{colorc}{HTML}{1476b5}
\colorpalette{colorc}
\definecolor{colorscorep}{rgb}{0,0.235,0.416}
\colorpalette{colorscorep}
\definecolor{colorfile}{rgb}{0.64,0.44.0}
\colorpalette{colorfile}
\colorlet{highlight}{red!60!black}
\definecolor{goldenhighlight}{HTML}{ffdb57} 
\colorlet{colorno}{red!80!black}
\colorlet{coloryes}{green!80!black}

\usepackage{pifont}
\newcommand{\cmark}{\textcolor{coloryes}{\ding{51}}} 
\newcommand{\xmark}{\textcolor{colorno}{\ding{55}}} 

\usepackage{listings}
\newcommand{\inputlisting}[4]{\lstinputlisting[language={#1}, float=ht, frame=tb, caption={#2}, label={#3}]{listings/#4}}
\newcommand{\fortranlisting}[3]{\inputlisting{[08]Fortran}{#1}{#2}{#3}}
\newcommand{\clisting}[3]{\inputlisting{C}{#1}{#2}{#3}}

\usepackage{tikz}
\usetikzlibrary{positioning} 
\usetikzlibrary{quotes} 
\usetikzlibrary{shapes.multipart} 
\usetikzlibrary{calc} 
\usetikzlibrary{decorations.pathreplacing} 
\usetikzlibrary{fit} 

\newlength{\letterheight}
\def\mpirelease#1{%
    {\includegraphics[height=\letterheight]{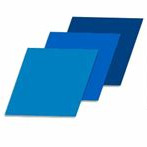} #1}
}

\tikzstyle{mpiversion} = [rectangle, rounded corners, fill=white]
\tikzstyle{fortranversion} = [rectangle, rounded corners, fill=colorfll]
\tikzstyle{scorepversion} = [rectangle, rounded corners, fill=colorscorepll]

\tikzstyle{programbox} = [rectangle, rounded corners, fill=black!5]
\tikzstyle{group} = [rectangle, rounded corners=0.5cm, draw, dashed, inner sep=0.25cm]

\tikzstyle{mpibox} = [rectangle, rounded corners, fill=white]
\tikzstyle{databox} = [rectangle, rounded corners, fill=colorfilelll]
\tikzstyle{providedby} = [draw, thick, <-]

\tikzset{%
    fileicon/.pic={
        \filldraw[sharp corners, fill=colorfilel] (0,\letterheight) node[coordinate](fileiconNW){} -- ($(0.625\letterheight, \letterheight) - (0.15\letterheight, 0) $) node[coordinate](fileiconNET){} -- ($(0.625\letterheight, \letterheight) - (0, 0.15\letterheight)$)  node[coordinate](fileiconNER){} [rounded corners=0.05\letterheight] -- (0.625\letterheight, 0) node[coordinate](fileiconSE){} -- (0,0) node[coordinate](fileiconSW){} -- cycle;
        \draw[sharp corners] (fileiconNET) |- (fileiconNER);
        \foreach \k in {1,2,3,4}
        {
            \draw[sharp corners, line cap=round] (fileiconSW) ++ ($ ({0.1*0.625*\letterheight}, 0) $) ++ ($ (0,{0.16*\k*\letterheight}) $) -- ++($ ({0.8*0.625*\letterheight}, 0) $);
        }
    }
}

\tikzstyle{procedure} = [draw, rectangle, rounded corners, align=center]
\tikzstyle{c-wrapper} = [procedure, fill=colorcl]
\tikzstyle{f-wrapper} = [procedure, fill=colorfl]
\tikzstyle{c-user} = [procedure, fill=colorcll]
\tikzstyle{f-user} = [procedure, fill=colorfll]
\tikzstyle{c-tool} = [procedure, draw=colorc, fill=white]
\tikzstyle{f-tool} = [procedure, draw=colorf, fill=white]
\tikzstyle{tool-impl} = [text=goldenhighlight]
\tikzstyle{mpi-sym} = [procedure, fill=white]
\tikzstyle{helper-node} = [opacity=0, align=center]
\tikzstyle{wraps} = [draw, thick, <->]
\tikzstyle{calls} = [draw, thick, dashed, ->]
\tikzstyle{calledby} = [draw, thick, dashed, <-]
\tikzstyle{wrapper-pic} = [
node distance=0.5cm,
every edge/.style={wraps},
every edge quotes/.style={auto, font=\footnotesize},
every label/.style={font=\footnotesize}]

\tikzstyle{pre-post-wrapper} = [procedure, rectangle split, rectangle split parts=3]
\tikzstyle{pre-post-c} = [pre-post-wrapper, rectangle split part fill={colorcl,white,colorcl}]
\tikzstyle{pre-post-f} = [pre-post-wrapper, rectangle split part fill={colorfl,white,colorfl}]

\tikzstyle{gearline} = [thick]
\newcommand{\gear}[4][]{%
    \def\modu{#2}
    \def\Zb{#3}
    \def\AngleA{#4}

    \pgfmathsetmacro{\Rpr}{\Zb*\modu/2}

    \pgfmathsetmacro{\Rb}{\Rpr*cos(\AngleA)}
    \pgfmathsetmacro{\Rt}{\Rpr+\modu}
    \pgfmathsetmacro{\Rp}{\Rpr-1.25*\modu}
    \pgfmathsetmacro{\AngleT}{sqrt(\Rt^2/\Rb^2-1)}

    \pgfmathsetmacro{\AnglePr}{180/pi*sqrt(\Rpr^2/\Rb^2-1)}
    \pgfmathsetmacro{\demiAngle}{180/\Zb}
    \pgfmathsetmacro{\Angledecal}{(\demiAngle+0.075*\AnglePr)/2}%

    \def\xxt{\Rb*(cos(\t r)+\t*sin(\t r))}
    \def\yyt{\Rb*(sin(\t r) - \t*cos(\t r))}

    \foreach \zz in{1,2,...,\Zb}{
        \coordinate(e\zz) at (\zz/\Zb*360+\Angledecal:\Rb);
        \draw[domain=-0:\AngleT,smooth,variable=\t,gearline,#1]
        plot ({atan2(\xxt,\yyt)-90+\zz/\Zb*360+\Angledecal}:{\Rb*sqrt(1+\t^2)}  )coordinate(f\zz);

        \coordinate(g\zz) at ({(\zz+1))/\Zb*360-\Angledecal}:\Rb);
        \draw[domain=-0:\AngleT,smooth,variable=\t,gearline,#1]
        plot ({atan2(\xxt,-\yyt)-90+(\zz)/\Zb*360-\Angledecal}:{\Rb*sqrt(1+\t^2)}  )coordinate(d\zz);

        \draw[gearline,#1] (f\zz) to[bend left=\demiAngle] (d\zz);

        \draw[gearline,rounded corners=\modu,#1](e\zz)  -- (\zz/\Zb*360+\Angledecal:\Rp) to[bend right=\demiAngle]  ({(\zz+1)/\Zb*360-\Angledecal}:\Rp)  -- (g\zz);
    }
    \draw[gearline, #1] (0,0) circle(0.2*\Rb);
}
\tikzset{%
    gearsicon/.pic={\gear[rounded corners=0pt]{0.2}{15}{20}
        \begin{scope}[xshift=2.8cm, rotate=15]
            \gear[rounded corners=0pt]{0.2}{12}{20}
        \end{scope}
}}
\tikzset{%
    gearsiconsmall/.pic={\gear[rounded corners=0pt, very thin]{0.2}{15}{20}
        \begin{scope}[xshift=2.85cm, rotate=15]
            \gear[rounded corners=0pt, very thin]{0.2}{12}{20}
        \end{scope}
}}

\newcommand{\code}[1]{\texttt{#1}}
\newcommand{\fts}{TS29113}

\newcommand{\todo}[1][]{}

\newcommand{\myabstract}{%
    Version 3.0 of the Message-Passing Interface (MPI) standard, released in 2012, introduced a new set of language bindings for Fortran 2008. By making use of modern language features and the enhanced interoperability with C, there was finally a type safe and standard conforming method to call MPI from Fortran. This highly recommended \texttt{use mpi\_f08} language binding has since then been widely adopted among developers of modern Fortran applications. However, tool support for the F08 bindings is still lacking almost a decade later, forcing users to recede to the less safe and convenient interfaces.
    Full support for the F08 bindings was added to the performance measurement infrastructure Score-P by implementing MPI wrappers in Fortran. Wrappers cover the latest MPI standard version 4.1 in its entirety, matching the features of the C wrappers. By implementing the wrappers in modern Fortran, we can provide full support for MPI procedures passing attributes, info objects, or callbacks. The implementation is regularly tested under the MPICH test suite. The new F08 wrappers were already used by two fluid dynamics simulation codes---Neko, a spectral finite-element code derived from Nek5000, and EPIC (Elliptical Parcel-In-Cell)---to successfully generate performance measurements.
    In this work, we additionally present our design considerations and sketch out the implementation, discussing the challenges we faced in the process. The key component of the implementation is a code generator that produces approximately 50k lines of MPI wrapper code to be used by Score-P, relying on the Python \texttt{pympistandard} module to provide programmatic access to the extracted data from the MPI standard.
}

\begin{document}

    \title{Performance measurements of modern Fortran MPI applications with Score-P}
    \date{\scriptsize Presented at the 15th International Parallel Tools Workshop 2024\footnote{https://tu-dresden.de/zih/das-department/termine/parallel-tools-workshop-2024}}
    \author{Gregor Corbin\footnote{Forschungszentrum Jülich GmbH, Jülich Supercomputing Center, \texttt{g.corbin@fz-juelich.de}}}

    \maketitle

    \section*{Abstract}
    \myabstract

    \section{Introduction}
    The Message Passing Interface (MPI) \cite{mpi41} is a community standard for distributed-memory parallelization, driven by contributors from science and industry.
    It is an integral part of many HPC applications.
    Started in the 1990s' it is being continuously extended and improved and currently in version 4.1.
    Due to its long history, the requirement to support a large variety of applications, and the choice to be backwards-compatible as much as possible, the MPI standard is a large document, describing over 400 procedures on over 1100 pages.

    Although its semantics are in principle independent of the programming language, MPI defines bindings for the C and Fortran programming languages.
    While the C bindings were always conforming with the ISO C standard, implementations of the Fortran bindings had to rely on non-standard extensions of Fortran, due to the limitations of the language at that time.
    For instance, buffers of arbitrary type can be passed as \code{void*} in C.
    Fortran until \fts{} (Further Interoperability with C) had no means to pass arbitrary types to a procedure.
    Implementations had to rely on unsafe implicit interfaces or non-standard compiler extensions to ignore type checking.

    Newer features of the Fortran language \cite{reid2014new} made it possible to define Fortran bindings for MPI which can be implemented conforming to the Fortran language.
    The technical specification \fts{} was introduced to that end \cite{fortran-ts29113}.
    For instance, choice buffers can be declared as \code{type(*), dimension(..)}.
    To accommodate the newer features, the entirely new Fortran 2008 language bindings have been introduced into the MPI standard.
    These bindings are available with the \code{mpi\_f08} module from MPI 3.0 onwards (released in 2012).
    The \code{mpi\_f08} module is since then the only recommended way to use MPI in Fortran.

    While the new language bindings have been part of MPI for over a decade, tool support is still lacking.
    We are not aware of any tool that supports the Fortran 2008 bindings entirely.
    The performance measurement tool Score-P \cite{scorep} in its currently released version 8.4 is no exception here.
    Developers are forced to recede to the unsafe and inconvenient Fortran 90 interface to enable the use of tools.

    Tools intercept MPI calls by providing wrappers that internally call the MPI library via the PMPI interface.
    Because MPI defines over 400 procedures, the wrappers are usually generated by a program.
    Since this is a common problem for tools, multiple such wrapper generators have been developed.
    LLNL wrap.py \cite{wrappy} fills out user-defined templates with information from \code{mpi.h} to produce wrappers.
    This is ideal for lightweight wrappers that act similarly for each function.
    Because it relies solely on C headers, it cannot be used to produce wrappers for the Fortran 2008 bindings.
    WMPI \cite{wmpi} provides a hierarchy of wrapper layers that funnel calls from C and Fortran into a single tool layer written in C.
    Using this infrastructure, a tool developer only has to provide wrappers for the C bindings.
    The paper is an excellent reference for the many issues (see Section \ref{sec:incompatibilities}) tools have to solve when dealing with Fortran user codes.
    Since both mentioned projects have not been maintained for several years we decided to develop our own wrapper generator.
    Last but not least, the MPI implementors themselves face similar issues as tools developers.
    Zhang et al. \cite{zhang2014implementing} discuss how the Fortran 2008 bindings are implemented in MPICH \cite{mpichgit}.
    In contrast to an external tool, they can focus on their own MPI library and have control over the internals of their library.

    Score-P \cite{scorep} is a highly scalable tool for profiling, i.e.,\ summarizing program execution,  and event tracing, i.e.,\ capturing events in chronological order, of HPC applications.
    Score-P adds instrumentation hooks into a user's application by either prepending or replacing the compile and link commands.
    C, C++, Fortran, and Python codes as well as many HPC programming models (MPI, threading, GPUs, I/O) are supported.
    Together with analysis tools build on top of its output formats, Score-P provides insight into massively parallel HPC applications, their communication, synchronization, I/O, and scaling behavior allowing HPC users to pinpoint performance bottlenecks and their causes.

    We added full support for the Fortran 2008 bindings to Score-P, by implementing MPI wrappers in Fortran.
    These wrappers cover the MPI 4.1 standard entirely and match Score-P's C wrappers in features.
    The wrappers are generated by a Python program that builds upon the pympistandard \cite{pympistandard} tool to access MPI procedure signatures.
    We support the two major MPI implementations OpenMPI \cite{openmpi} and MPICH \cite{mpichgit}.

    The remainder of this work is organized as follows.
    Section \ref{sec:old-vs-new} highlights the advantages of the Fortran 2008 bindings over the Fortran bindings.
    In Section \ref{sec:tools-and-mpi} we recap how tools can intercept calls to the MPI library, and summarize the issues associated with a mixed-language environment.
    Then, in Section \ref{sec:fortran-wrappers-in-scorep} we discuss the design of the new Fortran 2008 wrappers in Score-P in contrast to the status quo.
    A major part of the work was the design and implementation of the program that generates the wrapper code.
    We present this tool in Section \ref{sec:wrapper-generator}.
    Finally, Section \ref{sec:conclusion} contains a summary of the work and some concluding remarks.

    \section{A case for using the Fortran 2008 bindings}
    \label{sec:old-vs-new}
    Given the drastic evolution between Fortran 77 and Fortran 2008 during the history of MPI, Fortran support for MPI is a complicated matter.
    The MPI standard \cite[Ch. 19]{mpi41} devotes an entire chapter of more than 60 pages to this topic.
    MPI defines two sets of language bindings for Fortran: the Fortran bindings and the Fortran 2008 bindings.

    The Fortran bindings are designed to work with Fortran 90, or even Fortran 77 which restricts the use of language features.
    The same binding is available via two so-called support methods.
    Fortran 77 codes can only access MPI by including the \code{mpif.h} file, as this version of the language does not have modules and explicit interfaces.
    All interfaces are implicit, thus there is no argument checking performed by the compiler.
    From Fortran 90 on, it is possible to use the \code{mpi} module instead.
    While this allows argument checking in principle, it is severely limited in practice.
    First, all MPI handles are integers, which makes it easy to accidentally pass the wrong arguments.
    Second, some MPI implementations, e.g.\ MPICH, fall back to the Fortran 77 implementation to disable argument checking for routines accepting choice buffers \cite{zhang2014implementing}.

    In contrast, the Fortran 2008 bindings employ many of the newer language features to provide a safer and more convenient interface.
    These bindings are exclusively available by using the \code{mpi\_f08} module.
    Only this support method is recommended in the MPI standard and fully compliant with the Fortran language\footnote{Full compliance needs \fts{}.}.
    Additionally, the \code{mpi\_f08} module provides these advantages:
    \begin{itemize}
        \item The compiler can check all arguments.
        \item MPI handles have their own types, e.g.\ \code{type(MPI\_Comm)}.
        \item Non-contiguous buffers can be passed to non-blocking MPI routines\footnote{If MPI sets \code{MPI\_SUBARRAYS\_SUPPORTED} to \code{.true.}.}.
        \item Buffers of non-blocking operations can be protected by the \code{ASYNCHRONOUS} attribute\footnote{If MPI sets \code{MPI\_ASYNC\_PROTECTS\_NONBLOCKING} to \code{.true.}.}.
        \item Large-count overloads, also known as embiggened procedures, are available.
        \item The \code{ierror} argument is optional\footnote{Optional arguments are a language feature since Fortran 90. In practice, one should always include the argument with the older support methods.}.
    \end{itemize}

    Listings \ref{lst:mpi-f08-example} and \ref{lst:mpi-example} show example implementations of a two-dimensional halo exchange using the Fortran 2008 bindings, and the Fortran bindings, respectively.
    A comparison between the two examples demonstrates the mentioned benefits of the Fortran 2008 bindings.

    \fortranlisting{Fortran 2008 example with \code{use mpi\_f08}}{lst:mpi-f08-example}{halo-exchange-f08.F90}
    \fortranlisting{Fortran example with \code{use mpi}}{lst:mpi-example}{halo-exchange.F90}

    \section{Tools and MPI}
    \label{sec:tools-and-mpi}

    \subsection{PMPI interface and library interposition}
    MPI defines an interface which allows external tools to intercept calls to the MPI library \cite[Ch. 15.2, pp. 717]{mpi41}.
    Each MPI procedure is exposed under a different name in this interface, starting with \code{PMPI\_} instead of \code{MPI\_}.
    A tool that intercepts \code{MPI\_Send} for instance, provides a wrapper that delegates the MPI functionality to \code{PMPI\_Send}.
    In addition, the wrapper can do tool-specific work, for example recording time stamps and message sizes.

    To intercept a call to MPI, the tool links a symbol to the application that overrides the same symbol provided by the MPI library.
    In C, the symbol name is identical to the procedure name.
    But symbol names for Fortran bindings are more complex \cite[Ch. 19.1.5]{mpi41}.
    The library provides one symbol for each supported binding (Fortran/Fortran 2008).
    Additionally, the symbol name indicates whether the routine passes choice buffers with array descriptors (\code{\_fts} or \code{\_f08ts} suffix), and whether the symbol is the large-count overload (\code{\_c} suffix).
    Finally, the compiler mangles the symbol name.
    The usual mangling schemes convert the name to lowercase or uppercase and append one or two underscores.
    To intercept Fortran calls, the tool has to either provide symbols for all these variants, or determine which symbol names are actually present in the MPI library.

    \subsection{From Fortran user code to C tool code}
    \label{sec:incompatibilities}
    Score-P is written in C, thus it is straightforward to intercept MPI calls that come from C.
    But due to intrinsic differences between the two languages and their respective MPI bindings, the situation is more complicated when a call from Fortran is intercepted by a tool written in C.
    In the following we summarize the issues that any such tool has to address.
    Some are relevant when evaluating MPI arguments in the tool, some are relevant when delegating to the PMPI call in the wrapper, and some are relevant in both cases.
    Most have been discussed previously, see e.g. \cite{wmpi} for items 1 to 8, and \cite{zhang2014implementing} for 1,4-9 along with possible solutions.
    \begin{enumerate}
        \item \textbf{Fortran logical: }Fortran has an intrinsic \code{logical} type, whose internal representations for \code{.true.} and \code{.false.} do not have to match with C. Fortran \code{logical}s have to be converted to \code{integer}, or \code{logical(kind=c\_bool)} available in Fortran 2003.
        \item \textbf{Error return type: }In Fortran, the MPI error code is returned by an additional argument \code{integer, intent(out) :: ierr}. This argument is mandatory in the Fortran bindings, but optional in the Fortran 2008 bindings. Whether the error argument is actually present has to be checked in Fortran.
        \item \textbf{Fortran only routines: }A few MPI procedures (e.g.\ \code{MPI\_F\_sync\_reg}) exist only in the Fortran bindings. Wrappers for these procedures cannot delegate to the C PMPI function.
        \item \textbf{Callbacks, Attributes, Choice buffers: }Wrappers for routines that have callback arguments, choice buffer arguments, or attribute caching routines, must call the matching PMPI function in the same language and support method \cite[Ch. 19.1.5]{mpi41}.
        \item \textbf{Array descriptors: }The Fortran and Fortran 2008 bindings allow two methods to pass choice buffers. Buffers can be passed by address, which translates into a \code{void*} argument in C. If supported by the compiler, buffers can be passed by array descriptor, using the \code{type(*), dimension(..)} syntax. This calling convention is encoded into the specific procedure name. Routines passing array descriptors are marked with the suffix \code{\_fts} (Fortran bindings) or \code{\_f08ts} (Fortran 2008 bindings).
        The intercepting routine has to use the same calling convention as the original call.
        Passing array descriptors to C is also not supported by all C compilers.
        \item \textbf{MPI handles: }In C, MPI handles are represented as opaque types, e.g.\ \code{MPI\_Comm} for communicator handles. In Fortran, handles are integers and in Fortran 2008, handles are \code{bind(c)} derived types that contain a single integer value. To convert between C and Fortran representations, MPI defines the \code{MPI\_Comm\_f2c} and \code{MPI\_Comm\_c2f} procedures, which are available exclusively in C.
        \item \textbf{MPI constants: }Some constants, e.g.\ \code{MPI\_BOTTOM}, \code{MPI\_STATUS\_IGNORE}, have different values in Fortran and C. When passing between the languages, one has to convert these values. Additionally, checking whether an argument is equal to one of these special constants is only possible in C.
        \item \textbf{Character strings: }Strings are pointers to null terminated character sequences in C, and fixed-length character arrays in Fortran. A conversion routine is needed to pass strings from Fortran to C or vice versa.
        \item \textbf{MPI Status object: }Status is represented as an array of integers of length \code{MPI\_STATUS\_SIZE} in Fortran, as the opaque type \code{type(MPI\_Status)} in Fortran 2008, and as the opaque type \code{MPI\_Status} in C.
        Similar to the handle types, the MPI standard defines calls to convert status objects between all three representations. But the calls converting to and from the Fortran 2008 representation are not provided by all MPI implementations, e.g.\ OpenMPI 4.0. Therefore, a tool cannot rely upon them. Writing custom conversion routines is also not possible, as status is an opaque type. A possible solution,  discussed in Section \ref{sec:scorep-fortran2008-wrappers}, is to pass language information along with the object and query all status properties in the original language.
        \item \textbf{Array indices: }Arrays indices start at zero in C, and at one in Fortran. Procedures taking array indices as arguments, e.g. \code{MPI\_Waitany}, use the numbering scheme of the calling language.
        \item \textbf{Info keys/values: }In the Fortran bindings, leading and trailing spaces are stripped from info arguments. In C, no such conversion is done. A tool might observe different values depending on the origin of the call. Passing an info key originating in Fortran to the C PMPI routine can change program behavior.
    \end{enumerate}

%
%

    \section{Fortran wrappers in Score-P}
    \label{sec:fortran-wrappers-in-scorep}
    In this section we summarize the state of MPI wrappers in Score-P before this work, which includes wrappers for C and Fortran 90.
    Then we discuss the design of the wrappers for Fortran 2008.

    As seen in Table \ref{tab:f2c-interface-support}, the Fortran wrappers do not handle all issues from Section \ref{sec:incompatibilities} correctly.
    In contrast, the new Fortran 2008 wrappers treat all listed issues correctly.
    More detailed discussions follow in Sections \ref{sec:scorep-fortran-wrappers} and \ref{sec:scorep-fortran2008-wrappers}.

    \begin{table}[h]
        \centering
        \begin{tabular}{rlcc}
            \toprule
            & Issue                        & \multicolumn{2}{c}{Correctly handled in wrapper layer} \\
            &                              & Fortran & Fortran 2008 \\
            \midrule
            1  & Logical                      & \xmark & \cmark \\
            2  & Error return                 & \cmark & \cmark \\
            3  & Fortran only routines        & \xmark & \cmark \\
            4  & Callbacks, Attributes,\dots  & \xmark & \cmark \\
            5  & Array descriptors            & \xmark & \cmark \\
            6  & MPI handles                  & \cmark & \cmark \\
            7  & MPI constants                & \cmark & \cmark \\
            8  & Character strings            & \cmark & \cmark \\
            9  & Status object                & \cmark & \cmark \\
            10 & Array indices                & \cmark & \cmark \\
            11 & Info strings                 & \xmark & \cmark \\
            \bottomrule
        \end{tabular}
        \caption{Summary of support in the Fortran and Fortran 2008 wrappers for the C/Fortran interface problems}
        \label{tab:f2c-interface-support}
    \end{table}

    \subsection{Design goals for Score-P's MPI wrappers}
    Score-P is widely used in the HPC community and installed on numerous HPC systems around the world.\todo{Examples with references}
    Therefore, portability is a primary consideration.
    We also aim to impose as little restrictions as possible on the user code.
    This means we want to support the two major MPI libraries, OpenMPI \cite{openmpi} and MPICH \cite{mpichgit}, in as many versions as possible.

    Unfortunately, no MPI library is bug free: Procedure signatures and symbol names may deviate from the standard, procedures may be declared in the header but not included in the library.
    Even if a bug is fixed in a newer version of the library, the old version may be in use for a long time.
    Additionally, users might rely on MPI 1 functions that have been removed from the standard but are still available in the library.
    Consequently, in Score-P we provide wrappers for all MPI procedures up to the current standard.
    During configuration, we detect which functions are provided by the MPI library and exclude all others.
    We also check for known variants in signatures.

    The wrappers should also be maintainable, i.e., it should be easy to add support for new features.
    This requirement is paramount in the design of the generator, see Section \ref{sec:wrapper-generator}, but also extends to the generated wrappers.
    A clear interface between the wrappers and the tool is advantageous.

    Run-time efficiency, while overall important for a measurement tool, is mostly relevant in the internal tool code and less of a consideration in the wrappers.
    We assume that in most practical cases an additional function call is negligible compared to the work that the tool does internally, and that Fortran to C conversions are cheap.

    Last but not least, Score-P already has wrappers for the C bindings and the Fortran 90 bindings.
    To avoid introducing new bugs, we do not modify the existing wrappers.
    Therefore, the Fortran 2008 layer exists parallel to the existing wrappers.
    Some changes to the internal interface from wrappers to the measurement system were necessary to support the Fortran 2008 wrappers.

    \subsection{The status quo of Fortran wrappers}
    \label{sec:scorep-fortran-wrappers}
\begin{figure}[h]
        \centering
        \begin{tikzpicture}[wrapper-pic]
            \node[c-user](cuser){\code{MPI\_Send();}};
            \node[f-user, right=5em of cuser, minimum width=12em](fuser){\code{call mpi\_send()}};

            \node[c-wrapper, below=of fuser.195](fwrapper){Intercept \code{mpi\_send\_}\\Marshall arguments} edge(fuser.195);
            \node[c-wrapper, tool-impl] at (cuser |- fwrapper)(cwrapper){Wrap \code{MPI\_Send}} edge(cuser) edge(fwrapper);

            \node[c-tool, left=of cwrapper]{Tool} edge[calledby](cwrapper);

            \node[helper-node, below=of fwrapper] (helper) {p\\p\\p};
            \node[mpi-sym] at(fuser.345 |- helper)(f08pmpi){\code{pmpi\_send\_fts\_}\\\code{pmpi\_send\_f08\_}\\\code{pmpi\_send\_f08ts\_}} edge(fuser.south -| f08pmpi);

            \node[mpi-sym, left=of f08pmpi](fpmpi){\code{pmpi\_send\_}};
            \node[mpi-sym] at(cwrapper |- fpmpi)(cpmpi){\code{PMPI\_Send}} edge(cwrapper);
        \end{tikzpicture}
        \caption{Architecture of MPI wrappers in the current release of Score-P (Version 8.4).}
        \label{fig:f-wrappers}
    \end{figure}
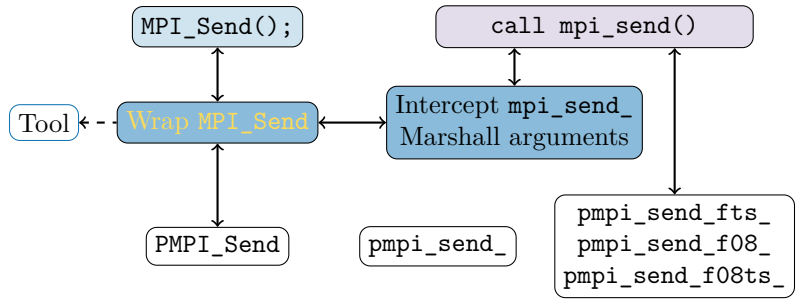

    Figure \ref{fig:f-wrappers} shows the architecture of the established C and Fortran wrappers in Score-P by example of \code{MPI\_Send}.

    On the one hand, a call to \code{MPI\_Send} from a C program is intercepted by the corresponding wrapper, written in C.
    This wrapper implements all functionality needed by Score-P, for instance recording entry and exit time stamps, and the number of bytes received.
    Therefore, it calls functions inside the tool, also written in C.
    The MPI function is completed by a call to the C symbol for \code{PMPI\_Send}.
    Arguments are mostly forwarded directly.
    On the other hand, a call to \code{mpi\_send} from a Fortran program is intercepted in a separate layer.
    This wrapper is written in C.
    Without the \code{bind(c)} interface, calling C functions from Fortran works by observing naming and argument type conventions.
    The naming convention is decided by the compiler and is determined while configuring Score-P.
    In the following, we represent the Fortran symbols by the lowercase procedure name with an underscore appended, which corresponds to a prominent mangling scheme.
    Listing \ref{lst:pmpi-send-f2c} shows an example wrapper implementation for \code{MPI\_Send}.
    Because Fortran passes arguments by reference, the arguments are declared pointers in C.
    MPI provides the \code{MPI\_Fint} type which is guaranteed to match a Fortran standard \code{integer}.
    The GCC website \cite{gcc-interop} presents more details on interoperability.

    The intercepting layer just converts the arguments from a Fortran representation to a C representation and then delegates to our C wrappers.
    We intercept only the calls from the Fortran bindings where buffers are passed by address (\code{mpi\_send\_}).
    These calls are redirected to use the C PMPI symbols (\code{PMPI\_Send}) internally, while the corresponding Fortran symbol (\code{pmpi\_send\_}) is never invoked.
    Symbols for the Fortran 2008 binding, or for passing buffers by array descriptor, are not intercepted.

    \clisting{Wrapper intercepting the Fortran \code{MPI\_Send} in C}{lst:pmpi-send-f2c}{pmpi-send-f2c.c}

    With respect to the issues listed in Section \ref{sec:incompatibilities}, this design has some deficits, as listed in Table \ref{tab:f2c-interface-support}.
    Fortran \code{logical}s are passed directly to C, which is not correct, but works for most compilers (item 1). Fortran-only routines are not intercepted, as these cannot delegate to the PMPI symbol in C (item 3). Calls to procedures with callback  arguments, calls to attribute caching procedures, and calls to procedures with choice buffers are completed in C, in violation with the standard (item 4). Info arguments are passed directly to C (item 11), which can change program behavior.

    \subsection{The new Fortran 2008 wrappers}
    \label{sec:scorep-fortran2008-wrappers}
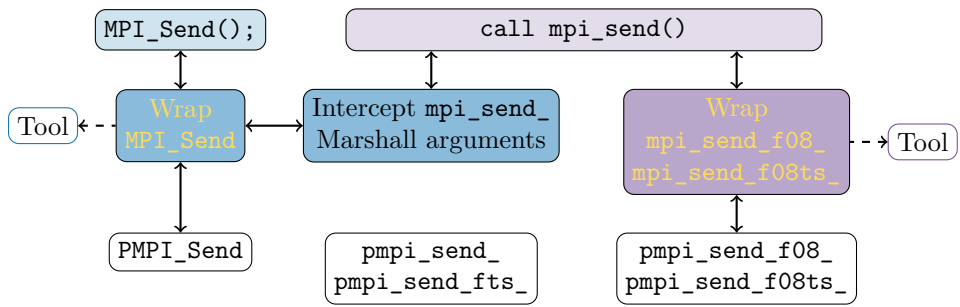
\begin{figure}[h]
        \centering
        \begin{tikzpicture}[wrapper-pic]
            \node[c-user](cuser){\code{MPI\_Send();}};
            \node[f-user, right=3em of cuser, minimum width=18em](fuser){\code{call mpi\_send()}};

            \node[c-wrapper, below=of fuser.188](fwrapper){Intercept \code{mpi\_send\_}\\Marshall arguments} edge(fuser.188);
            \node[f-wrapper, tool-impl, below=of fuser.352](f08wrapper){Wrap\\ \code{mpi\_send\_f08\_} \\ \code{mpi\_send\_f08ts\_}} edge(fuser.352);
            \node[c-wrapper, tool-impl] at (cuser |- fwrapper)(cwrapper){Wrap\\ \code{MPI\_Send}} edge(cuser) edge(fwrapper);

            \node[c-tool, left=of cwrapper](ctool){Tool} edge[calledby](cwrapper);
            \node[f-tool, right=of f08wrapper] {Tool} edge[calledby](f08wrapper);

            \node[mpi-sym, below=of f08wrapper] (f08pmpi){\code{pmpi\_send\_f08\_}\\ \code{pmpi\_send\_f08ts\_}} edge(f08wrapper);
            \node[mpi-sym] at (fwrapper |- f08pmpi)(fpmpi){\code{pmpi\_send\_}\\ \code{pmpi\_send\_fts\_}};
            \node[mpi-sym, anchor=north] at (cwrapper |- fpmpi.north)(cpmpi){\code{PMPI\_Send}} edge(cwrapper);
        \end{tikzpicture}
        \caption{Architecture of MPI wrappers in the upcoming release of Score-P, including the new Fortran 2008 wrappers.}
        \label{fig:f08-wrappers}
    \end{figure}

    In this section we discuss the design and implementation of the Fortran 2008 wrappers recently added to Score-P.
    Figure \ref{fig:f08-wrappers} shows the architecture of Score-P's MPI wrappers with addition of the new Fortran 2008 wrappers, using \code{MPI\_Send} as example.

    The Fortran 2008 wrappers intercept calls from applications that use the \code{mpi\_f08} module.
    These wrappers are separate from the other wrappers.
    Thus, applications relying on the Fortran and C bindings observe no change.

    We chose to implement these wrappers as a single layer of functions written in Fortran that call to the PMPI interface directly, closely mirroring the C wrappers.
    An example wrapper implementation for \code{MPI\_Send} is shown in Listing \ref{lst:f08-wrapper-example} in the following section.
    This design guarantees that the matching PMPI symbol is called, therefore avoids the issues around procedures taking callback, attribute or choice buffer arguments (item 4).
    We also do not need to convert the procedure arguments from Fortran to C and vice versa on the path to the PMPI call.
    MPI receives the arguments as they were provided by the user, unless deliberately modified by the wrapper.
    Consequently, the issues associated with Fortran to C conversion (items 1,5-10) are avoided there.

\begin{figure}[h]
        \centering
        \begin{tikzpicture}[wrapper-pic]
            \node[c-tool, minimum width=15em](ctool){\code{tool}};
            \node[c-tool, above=of ctool.10, label=0:C conversion] (fromf08){\code{tool\_fromf08}} edge(fromf08 |- ctool.north);

            \node[f-tool, above=of fromf08, label=0:Fortran conversion](ftool){\code{tool\_}} edge(fromf08);
            \node[f-wrapper, above=of ftool](f08wrapper){F08 wrapper\\ \code{call tool()}} edge[calls](ftool);
            \node[c-wrapper] at (ctool.170 |- f08wrapper)(cwrapper){C wrapper\\ \code{tool();}} edge[calls] (ctool.170);
        \end{tikzpicture}
        \caption{Calling Score-P functions from Fortran needs up to two additional function calls for conversion.}
        \label{fig:f08-tool-interface}
    \end{figure}
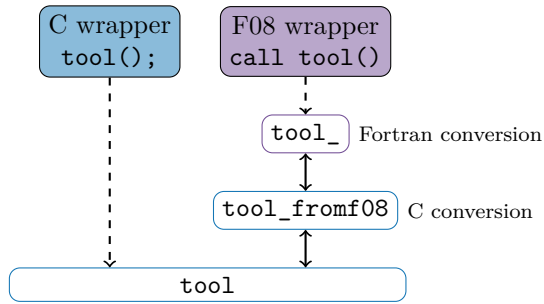

    Arguments passed to the tool still have to be converted to C.
    This is the responsibility of the Fortran interface to the tool, as shown in Figure \ref{fig:f08-tool-interface}.

    Calling a tool function from Fortran involves up to two additional interface layers.
    In the worst case, some parts of the conversion have to be done in Fortran, and other parts in C.
    Then, the call chain involves both conversion layers.
    In the best case, the function is interoperable, and the tool function is called directly from Fortran.
    On the one hand, we do only the necessary conversions with this design.
    On the other hand, the interface adds some maintenance costs.

    Finally, we discuss the additional complications when passing an \code{MPI\_Status} from Fortran to C.
    Although the MPI standard defines procedures to convert between a Fortran 2008 \code{type(MPI\_Status)} and a C \code{MPI\_Status}, not all MPI implementations provide these functions.
    Because the opaque status objects are implementation defined, we cannot implement these missing conversion functions.

\begin{figure}[h]
    \centering
    \begin{tikzpicture}[wrapper-pic,
        status-obj/.style={draw, rectangle split, rectangle split parts=2, font=\footnotesize}]

        \node[c-tool, minimum width=15em](ctool){\code{tool\_impl}};
        \node[c-tool, above=4em of ctool.10] (fromf08){\code{tool\_fromf08}} edge node[right=0.5em, status-obj]{\code{type(MPI\_Status) st}\nodepart{two}\code{lang=f08}} (fromf08 |- ctool.north);
        \node[f-tool, above=of fromf08](ftool){\code{tool\_}} edge(fromf08);
        \node[f-wrapper, above=of ftool](f08wrapper){F08 wrapper\\ \code{call tool()}} edge[calls](ftool);

        \node[c-tool, above=4em of ctool.170](cinterop){tool} edge node[left=0.5em, status-obj]{\code{MPI\_Status* st}\nodepart{two}\code{lang=c}} (ctool.170);
        \node[c-wrapper] at (cinterop |- f08wrapper)(cwrapper){C wrapper\\ \code{tool();}} edge[calls] (cinterop);

        \node[c-tool, below=of ctool, minimum width=12em](getcount){\code{get\_count}} edge[calledby](ctool);
        \node[mpi-sym, below=of getcount.190](cmpigetcount) {\code{MPI\_Get\_count}} edge["\code{lang==c}"](getcount.190);
        \node[f-tool, below=of getcount.350](fmpigetcounttof08) {\code{get\_count\_tof08}} edge["\code{lang==f08}"'](getcount.350);
        \node[mpi-sym, below=of fmpigetcounttof08](fmpigetcount) {\code{mpi\_get\_count\_}} edge(fmpigetcounttof08);
    \end{tikzpicture}
    \caption{Querying a status objects from C or Fortran in the tool requires additional work. A pointer to the object is passed along with a language tag. The tool queries the status object in the original language.}
    \label{fig:f08-status-interface}
    \end{figure}
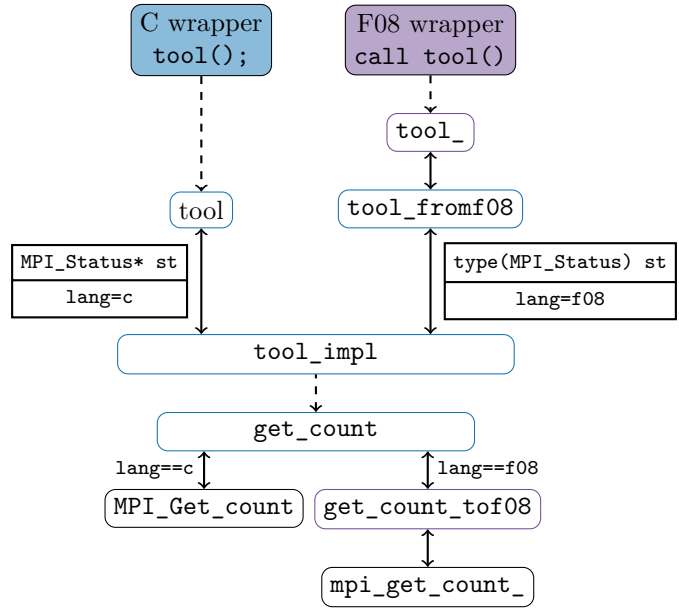

    Therefore, we pass a wrapper object, which contains a pointer to the status object and a language tag, to the tool.
    The tool function then queries the status object in its original language.
    Figure \ref{fig:f08-status-interface} depicts the necessary interface layers in this design.

    \section{The wrapper generating program}
    \label{sec:wrapper-generator}
    The MPI standard defines 491 procedures\footnote{Including removed interfaces, not counting large-count procedures.} in total, of which 393 have a Fortran 2008 binding\footnote{Handle conversion functions, \code{MPI\_T\_} functions and various other functions do not have a Fortran 2008 binding.}.
    Clearly, a program that writes wrappers for this many functions is beneficial.

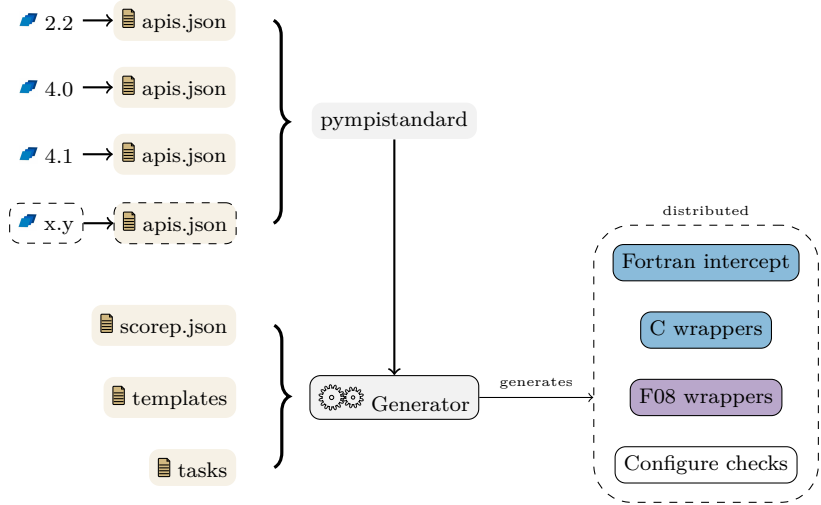
\begin{figure}[h]
        \centering
        \begin{tikzpicture}[node distance=0.4cm,
            font=\footnotesize,
            every edge quotes/.style={auto, font=\tiny},
            every label/.style={font=\tiny}]
            \node[mpibox] (mpi22) {\mpirelease{2.2}};
            \node[mpibox, below=of mpi22] (mpi40) {\mpirelease{4.0}};
            \node[mpibox, below=of mpi40] (mpi41) {\mpirelease{4.1}};
            \node[mpibox, draw, dashed, below=of mpi41] (mpixy) {\mpirelease{x.y}};

            \node[databox, right=of mpi22] (apis22) {\tikz{\pic {fileicon}} apis.json} edge[providedby] (mpi22);
            \node[databox, right=of mpi40] (apis40) {\tikz{\pic {fileicon}} apis.json} edge[providedby] (mpi40);
            \node[databox, right=of mpi41] (apis41) {\tikz{\pic {fileicon}} apis.json} edge[providedby] (mpi41);
            \node[databox, right=of mpixy, draw, dashed] (apisxy) {\tikz[solid]{\pic {fileicon}} apis.json} edge[providedby] (mpixy);

            \draw[decorate, very thick, decoration={brace, amplitude=0.2cm, raise=0.5cm}] (apis22.east) -- (apis22.east |- apisxy);

            \path (apis40.east) -- node[coordinate, midway](helper){} (apis41.east);
            \node[programbox, right=1cm of helper] (pympistandard) {pympistandard};
            \node[databox ,below=0.8cm of apisxy.south east, anchor=north east] (scorepjson) {\tikz{\pic {fileicon}} scorep.json};
            \node[databox, below=of scorepjson.south east, anchor=north east] (templates) {\tikz{\pic {fileicon}} templates};
            \node[databox, below=of templates.south east, anchor=north east] (tasks) {\tikz{\pic {fileicon}} tasks};
            \draw[decorate, very thick, decoration={brace, amplitude=0.2cm, raise=0.5cm}] (scorepjson.east) -- (tasks.east -| scorepjson.east);

            \path (pympistandard |- templates) node[programbox, draw] (driver) {\tikz{\pic[scale=0.1]{gearsiconsmall}} Generator};

            \draw[providedby] (driver) edge[<-] (pympistandard);

            \node[f-wrapper, right=2cm of driver](f08wrapper) {F08 wrappers};
            \node[c-wrapper, above= of f08wrapper](cwrapper) {C wrappers};
            \node[c-wrapper, above= of cwrapper](fwrapper) {Fortran intercept};
            \node[procedure, below=of f08wrapper](configurechecks) {Configure checks};
            \node[group, fit=(fwrapper) (cwrapper) (f08wrapper) (configurechecks), label=90:distributed](generated) {};
            \path (driver) edge[->, "generates"] (driver -| generated.west);
        \end{tikzpicture}
        \caption{Inputs and outputs of the wrapper generator.}
        \label{fig:generator}
    \end{figure}

    In this section we describe the  program we developed specifically for generating Score-P's MPI wrappers.
    Since the problem of wrapping MPI is common for tools, there exist other generator tools already, for example LLNL wrap \cite{wrappy}.
    In the following we also discuss the specific requirements and resulting design choices that led us to write our own generator.
    Figure \ref{fig:generator} presents a high-level overview of the generator.

    The first important consideration is the source of information on procedure interfaces.
    LLNL wrap and WMPI use C headers or Fortran module files provided by the MPI installation.
    We decided against that approach, because it means that the information is only available at the time of installation of Score-P, therefore requiring the generator tool to be distributed with Score-P.

    Instead, we rely on the machine-readable binding specification \code{apis.json} provided by the MPI standard since version 4.0.
    The package pympistandard \cite{pympistandard} allows convenient access to this information from Python.
    We merge the \code{apis.json} files for all versions of the MPI standard into one file which is then used as input by pympistandard.
    This allows for easy inclusion of future versions of the standard.
    Score-P also supports interfaces that were removed from MPI in version 3.3.
    But these removed interfaces are not included in any \code{apis.json} file.
    We provide a handwritten file for the these removed interfaces.

    Only the generated wrapper code is distributed with Score-P.
    This includes wrappers for the C, Fortran, and Fortran 2008 bindings.
    At the time of installation, the wrappers are adapted to the specific MPI library via configure checks.
    There is one check for each procedure in the Fortran 2008 bindings that determines whether the function is accessible in the \code{mpi\_f08} module, and under which specific symbol name it is present in the MPI library.

    Figure \ref{fig:generator} shows additional inputs for the generator.
    These files define which code is generated for each wrapper and the overall organization of wrappers into source files.

    Before we detail our approach to wrapper generation, some remarks on the wrapper code are in order.
    A sketch of the \code{MPI\_Send} wrapper in Listing \ref{lst:f08-wrapper-example} serves as example.
    All wrappers have the same common structure: First, a preprocessor guard to include/remove the wrapper at compilation time. Second, the function header with use statements, dummy argument declarations and local variable declarations. Third, the function body, which consists of the code before the PMPI call, the PMPI call and the code after. All wrappers check at runtime, whether they should write events to the trace and if so write at least an \code{ENTER} and an \code{EXIT} event.
    On top of that basic functionality, a wrapper might execute additional code according to the  semantics of the MPI call.
    Often, the extended functionality is similar for groups of wrappers.
    For instance, all functions that send a point-to-point message record the number of bytes sent.
    However, some parametrization might be necessary to account for differently named procedure arguments.
    Finally, some wrappers implement unique and specialized behavior, for instance \code{MPI\_Finalize}.

    \fortranlisting{Simplified schematic of Score-P wrappers by example of \code{MPI\_Send}}{lst:f08-wrapper-example}{wrapper-structure.F90}

\begin{figure}
        \centering
        \tikzstyle{component} = [draw, thick, align=left, rectangle split, rectangle split parts=2, rectangle split part fill={colorfilell, white}]
        \tikzstyle{hook} = [draw, rectangle, align=left, fill=black!5]
        \tikzstyle{render} = [rectangle, outer sep=0pt]
        \tikzstyle{flow} = [draw, thick, ->, rounded corners=2pt]
        \newcommand{\tab}{{\ \ }}
        \begin{tikzpicture}[remember picture, wrapper-pic, font=\scriptsize, node distance=0.5cm]
            \node[component] (task) {%
                Task\\
                "calc\_bytes\_sent"
                \nodepart{two}%
                \tikz[remember picture, node distance=2ex]{
                    \node[render] (renderlocalvars) {local\_vars()};
                    \node[render, below = of renderlocalvars.south west, anchor=north west] (renderenter) {render\_enter()};
                    \node[render, below = of renderenter.south west, anchor=north west ] (renderexit) {render\_exit()};}
            };
            \node[component, above=of task] (pympistandard) {%
                pympistandard
                \nodepart{two}
                "mpi\_send" : $\lbrace$ \\
                \tab "parameters: ... \\
                \tab "attributes: ...
            };
            \node[component, below=of task] (scorepjson) {%
                scorep.json%
                \nodepart{two}%
                "mpi\_send" : $\lbrace$\\
                \tab "tasks" : [ \\
                \tab \tab "calc\_bytes\_sent"\\
                \tab \tab ...
            };

            \node[component, right=of task] (skel) {%
                Wrapper\\
                "MPI\_Send"
                \nodepart{two}\ttfamily%
                \tikz[remember picture] {\node[rectangle,fill=none,inner sep=0](wrapperhead){subroutine MPI\_Send(..., count, ...)};}\\
                \tab \tikz[remember picture]{\node[hook](hooklocalvars) {integer :: bytes\_sent};}\\
                \tab if ( event\_gen\_active("MPI\_Send") ) then \\
                \tab\tab call write\_event("ENTER MPI\_Send") \\
                \tab\tab \tikz[remember picture]{\node[hook](hookenter) {bytes\_sent = count * ...};}\\
                \tab end if \\
                \tab call PMPI\_Send(...) \\
                \tab if ( event\_gen\_active("MPI\_Send")) then \\
                \tab\tab \tikz[remember picture]{\node[hook](hookexit) {\vphantom{b}...};}\\
                \tab\tab call write\_event("EXIT MPI\_Send") \\
                \tab end if\\
                end subroutine
            };
            \node[component, right=of skel] (template) {%
                Template\\"p2p.F90"
                \nodepart{two}%
                \tikz[remember picture]{\node(mpisend)[render, inner sep=0pt]{\code{MPI\_Send}};}\\
                \code{MPI\_Bsend}\\
                \dots%
            };

            \draw[flow, <-] (mpisend.west) ++(-1pt, 0pt) -- (skel.east |- mpisend);

            \path (skel.west) -- node[coordinate] (helper){} (task.east);
            \draw[flow] (renderlocalvars) -- (helper |- renderlocalvars) |- (hooklocalvars);
            \draw[flow] (renderenter) -- (helper |- renderenter) |- (hookenter);
            \draw[flow] (renderexit) -- (helper |- renderexit) |- (hookexit);

            \draw[flow] (pympistandard) -| (wrapperhead.172);

            \draw[flow] (scorepjson) -- (task);
            \draw[flow] (pympistandard) -- (task);
        \end{tikzpicture}
        \caption{Interplay of the various pieces of information in the code generation process.}
        \label{fig:generator-internals}
    \end{figure}
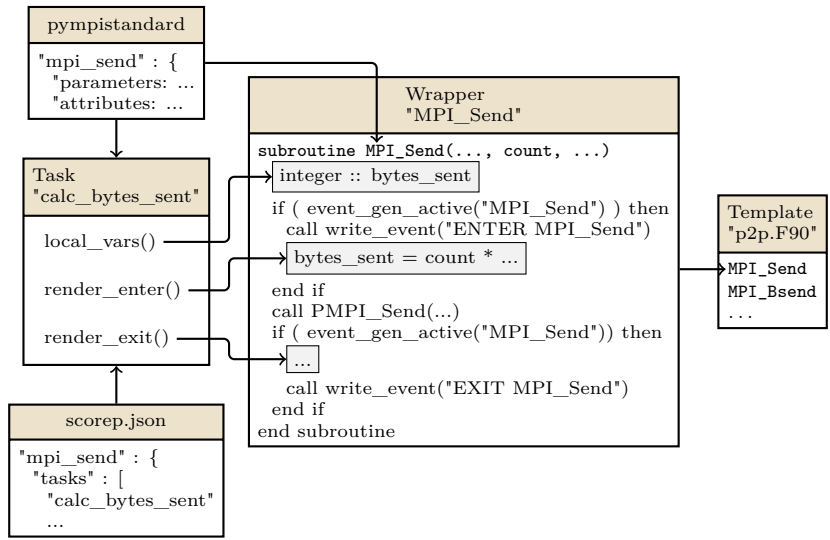

    With these observations in mind we discuss the code generation.
    Figure \ref{fig:generator-internals} provides a schematic of the process.
    The wrappers are organized in several source files, loosely corresponding to chapters in the MPI standard, e.g.\ point-to-point communication, collective communication, communicators and groups, and so on.
    Each source file is generated from a template which contains a list of wrappers to generate and potentially some common code.
    The template only defines which wrappers should be generated, not how this is done.

    The parametrized common structure is rendered by a python function that takes the MPI procedure name as input and uses the interface definition provided by pympistandard to fill in names and types of arguments.
    A wrapper can be extended by an arbitrary number of so-called tasks to extend its behavior.
    A task bundles a small piece of additional functionality exhibited by the wrapper.
    Which tasks are added to each wrapper is defined in the \code{scorep.json} file.
    Tasks are reusable, parametrized and orthogonal to each other.
    For instance, the \code{MPI\_Send} wrapper is extended by a task to calculate the number of bytes sent.
    The same task is reused by all other point-to-point sending procedures.
    The calculation depends on the count argument to the MPI procedure.
    Since the name of this argument varies between procedures, the task is parametrized in this regard.
    Tasks are mostly orthogonal and do not interfere with each other.
    The prime example is the wrapper for \code{MPI\_Sendrecv} which can be implemented by adding the tasks for \code{MPI\_Send} and \code{MPI\_Recv}.

    A task may add code to multiple places in the wrapper.
    For instance, to calculate the sent bytes a local variable is declared at the beginning of the wrapper, which is later set to the result of the calculation.
    Therefore, the rendering function defines hooks where the tasks can insert their code.
    Some hook points are shown in Listing \ref{lst:f08-wrapper-example}.

    \section{Conclusion}
    \label{sec:conclusion}
    With the newly added wrappers, Score-P is one of the first tools to offer support for Fortran codes that employ the modern and recommended Fortran 2008 bindings of MPI.
    Score-P's Fortran 2008 wrappers provide the same features as the C wrappers.
    The new wrappers are compatible with relatively recent versions of GCC\footnote{Tested with GCC 11}, Clang/Flang\footnote{Needs a development version}, Cray\footnote{Tested with PrgEnv-cray 6.0.10}, Intel\footnote{Tested with PrgEnv-intel 6.0.10}, and NVHPC\footnote{Tested with NVHPC 23.7}.
    The two major MPI implementations OpenMPI and MPICH, and derivatives such as ParaStationMPI are supported.
    The feature will be available in an upcoming release of Score-P\footnote{Please contact support@score-p.org for access to a development version.}.

    We developed a code generator to write wrappers automatically based on information on the MPI standard.
    This wrapper generator is tailored specifically to the requirements of Score-P.
    We emphasize maintainability and extensibility in the design of the generator, such that supporting new MPI procedures and adding features to the wrappers will be easy in the future.

    Only the generated wrappers are distributed with the release versions of Score-P.
    On the one hand, the generator does not add a dependency for the users.
    On the other hand, considerable effort has to be spent at installation to configure the wrappers for the user's toolchain.

    The new wrappers have been tested threefold.
    First, our CI verifies that Score-P builds successfully and can instrument and run basic test programs.
    This is done on about 100 different system/compiler/MPI combinations.
    Second, we run the MPICH test suite with Score-P as the compiler to verify that Score-P does not invalidate correct MPI programs.
    Third, we successfully instrumented and recorded traces for two application codes that use the Fortran 2008 bindings: Neko\cite{neko, nekogit} and EPIC \cite{epic, epicgit}.


    \bibliographystyle{spmpsci}
    \bibliography{literature.bib}
\end{document}